\documentclass[twocolumn]{el-author}

\newcommand{\ua}{\uparrow}
\newcommand{\nc}{\newcommand}
\nc{\da}{\downarrow} \nc{\hc}{\hat{c}} \nc{\hS}{\hat{S}}
\nc{\bra}{\langle} \nc{\ket}{\rangle} \nc{\eq}{equation (\ref}
\nc{\h}{\hat} \nc{\hT}{\h{T}}\nc{\be}{\begin{eqnarray}}
\nc{\ee}{\end{eqnarray}}\nc{\rd}{\textrm{d}}\nc{\e}{eqnarray}\nc{\hR}{\hat{R}}\nc{\Tr}{\mathrm{Tr}}
\nc{\tS}{\tilde{S}}\nc{\tr}{\mathrm{tr}}\nc{\8}{\infty}\nc{\lgs}{\bra\ua,\phi|}\nc{\rgs}{|\ua,\phi\ket}
\nc{\hU}{\hat{U}}\nc{\lfs}{\bra\phi|}\nc{\rfs}{|\phi\ket}\nc{\hZ}{\hat{Z}}\nc{\hd}{\hat{d}}\nc{\mD}{\mathcal{D}}
\nc{\bd}{\bar{d}}\nc{\bc}{\bar{c}}\nc{\mc}{\mathcal}\nc{\ea}{eqnarray}\nc{\mG}{\mathcal{G}}\nc{\bce}{\begin{center}}
\nc{\ece}{\end{center}}
\date{2018}

\begin{document}

\title{Set-membership NLMS algorithm based on bias-compensated and regression noise variance estimation for noisy inputs}

\author{Kaili Yin, Haiquan Zhao and Lu Lu}

\abstract{The bias-compensated set-membership normalised LMS (BCSM-NLMS) algorithm is proposed based on the concept of set-membership filtering, which incorporates the bias-compensation technique to mitigate the negative effect of noisy inputs. Moreover, an efficient regression noise variance estimation method is developed by taking the iterative-shrinkage method. Simulations in the context of system identification demonstrate that the misalignment of the proposed BCSM-NLMS algorithm is low for noisy inputs.}

\maketitle

\section{Introduction}

For system identification problem, the normalised least-mean-square (NLMS) algorithm is widely used due to its low computational complexity and robustness \cite{1}. By employing the set-membership filtering (SMF) approach, the set-membership NLMS (SM-NLMS) algorithm was proposed, which keeps the merits of the NLMS algorithm and reduces the computational complexity in practical applications \cite{2}. Moreover, many variants of the SMF-based algorithms were also proposed \cite{3,4}.

The SMF approach has good performance if the deleterious effect of the input noise is not taken into account. However, it is well known that such input noise can cause bias estimation in adaptive filtering. Under this scenario, the SM-NLMS algorithm may suffer from the large steady-state misalignment during adaptation. To eliminate the bias due to noisy inputs, the bias-compensated strategies were proposed in several previous studies \cite{5,6}. In these works and other similar references on this topic, the adaptation typically relies on the use of an unbiasedness criterion and less attention is paid to the SMF solution.

Motivated by these considerations, the bias-compensated set-membership NLMS (BCSM-NLMS) algorithm based on the unbiasedness criterion and the SMF method is proposed in this paper. Additionally, a new regression noise variance estimation method is proposed by using iterative-shrinkage scheme. Simulation results show the usefulness of the proposed algorithm for noisy input data. 

\section{SM-NLMS algorithm}

Consider a system identification model, the desired signal $d(n)$ can be expressed as
\begin{equation}
d(n) = {\bm x^{\mathrm T}}(n){\bm w_o} + v(n)
\label{001}
\end{equation}
where $\bm x(n)=[x(n),x(n-1),...,x(n-L+1)]^{\mathrm T}$ denotes the input vector, $[\cdot]^{\mathrm{T}}$ denotes the transposition operation, $\bm w_o$ is the unknown parameter vector with length of $L$, and $v(n)$ is a zero-mean white Gaussian noise (WGN) with variance $\sigma_v^2$. The update equation of the conventional SM-NLMS algorithm is expressed as \cite{2} 
\begin{equation}
{\bm{w}}(n+1) = \left\{ \begin{array}{l}
{\bm{w}}(n) + \mu \frac{{\bm x(n)}}{{{\bm x^{\mathrm T}}(n)\bm x(n)}}\left\{ {e(n) - \gamma} \right\}\;\;\;{\mathrm {if}}\left| {e(n)} \right| > \gamma  \\ 
{\bm{w}}(n)\;\;\;\;\;\;\;\;\;\;\;\;\;\;\;\;\;\;\;\;\;\;\;\;\;\;\;\;\;\;\;\;\;\;\;\;\;\;\;\;\;\;\;\;\;\;\;\;\;\;\;\;\;\;\;\;\;\;\;\;\; {\mathrm {otherwise}} \\ 
\end{array} \right.
\label{003}
\end{equation}
where $\bm w(n)$ is the weight vector of the adaptive filter, $\gamma$ is the error bound, $\mu$ is the step size, and $e(n)$ is the error signal, calculated by 
\begin{equation}
e(n) = d(n) - \bm x^{\mathrm T}(n)\bm{w}(n).
\label{b01}
\end{equation}

\section{Proposed BCSM-NLMS algorithm}
\begin{figure}[h]
	\centering
	\includegraphics[scale=0.55] {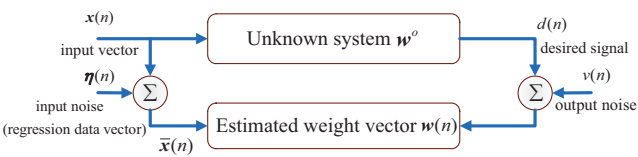}
	\caption{\emph{Measurement model with noisy inputs.}}
	\label{Fig01}
\end{figure}

Consider a measurement model (Fig. \ref{Fig01}).  The noisy input signal is expressed as
\begin{equation}
{\overline {\bm x}} (n) = \bm x(n) + \bm \eta(n)
\label{004}
\end{equation}
where $\bm \eta(n)=[\eta(n),\eta(n-1),...,\eta(n-L+1)]$ denotes the regression noise signal with zero-mean and variance $\sigma_\eta^2$.

To eliminate the bias caused by noisy input, a bias-compensation vector $\bm \xi(n)$ is introduced
\begin{equation}
\bm{w}(n+1) = \left\{\begin{array}{l}
\bm{w}(n) + \mu \frac{{\overline {\bm x} (n)}}{{\overline {\bm x}^{\mathrm T}}(n)\overline {\bm x}(n)}\left\{ \overline e(n) - \gamma \right\} + \bm \xi(n)\;\;\;{\mathrm {if}} \left| \overline e(n) \right| > \gamma  \\ 
\bm{w}(n) \;\;\;\;\;\;\;\;\;\;\;\;\;\;\;\;\;\;\;\;\;\;\;\;\;\;\;\;\;\;\;\;\;\;\;\;\;\;\;\;\;\;\;\;\;\;\;\;\;\;\;\;\;\;\;\;\;\;\;\;\;\;\;\;\;\;\;\;\;\;\;\;\;\;\;\;\; {\mathrm {otherwise}} \\ 
\end{array} \right.
\label{005}
\end{equation}
where $\overline e(n)$ is defined as
\begin{equation}
\overline e(n) \triangleq d(n) - \overline {\bm x}^{\mathrm T}(n)\bm{w}(n).
\label{b02}
\end{equation}

Define the weight deviation vector as follow:
\begin{equation}
\widetilde {\bm w}(n) \triangleq \bm w_o - \bm {w}(n).
\label{b03}
\end{equation} 
Thus, (\ref{005}) becomes
\begin{equation}
\widetilde {\bm w}(n+1) = \left\{\begin{array}{l}
\widetilde {\bm w}(n) - \mu \frac{\overline {\bm x}(n)}{\overline {\bm x}^{\mathrm T}(n)\overline {\bm x}(n)}\left\{ \overline e (n) - \gamma \right\} - \bm \xi(n)\;\;\;{\mathrm {if}} \left| {\overline e(n)} \right| > \gamma  \\ 
\widetilde {\bm w}(n)\;\;\;\;\;\;\;\;\;\;\;\;\;\;\;\;\;\;\;\;\;\;\;\;\;\;\;\;\;\;\;\;\;\;\;\;\;\;\;\;\;\;\;\;\;\;\;\;\;\;\;\;\;\;\;\;\;\;\;\;\;\;\;\;\;\;\;\;\;\;\;\;\;\;\;\;{\mathrm {otherwise.}}  \\ 
\end{array} \right.
\label{006}
\end{equation}

To obtain the unbiased estimation,  we consider using the following unbiasedness criterion: \cite{3}: 
\begin{equation}
\mathbb{E}\left[ {{\widetilde {\bm w}}(n+1)|{\overline {\bm x}}(n)} \right] = 0 \;\;{\mathrm {whenever}}\;\;\mathbb{E}\left[ {{\widetilde {\bm w}}(n)|{\overline {\bm x}}(n)} \right] = 0
\label{007}
\end{equation}
where $\mathbb{E}\left[\cdot\right]$ denotes the expectation of a random variable. Taking expectation on both sides of (\ref{006}) for $\left| {\overline e(n)} \right| > \gamma$ and considering (\ref{007}), we have
\begin{equation}
\begin{aligned}
&\mathbb{E}\left[ {\widetilde {\bm w}(n+1)|\overline {\bm x}(n)} \right] \\ 
 =&\; \mathbb{E}\left[ {\widetilde {\bm w}(n)|\overline {\bm x}(n)} \right] - \mu \mathbb{E}\left[ {\left.\frac{{{\overline {\bm x}}(n)}}{\overline {\bm x}^{\mathrm T}(n)\overline {\bm x}(n)}\left\{ {\overline e(n) - \gamma} \right\}  \right|\overline {\bm x}(n)} \right]  \\ 
&- \mathbb{E}\left[ {\bm \xi(n)|\overline {\bm x}(n)} \right] = 0.
\end{aligned}
\label{008}
\end{equation}

Taking the following equations into consideration:
\begin{equation}
\begin{aligned}
\overline e(n) =&\; d(n) - \overline {\bm x}^{\mathrm T}(n)\bm{w}(n) \\
=&\; e(n) - \bm \eta^{\mathrm T}(n)\bm{w}(n)
\label{009}
\end{aligned}
\end{equation}
\begin{equation}
e(n) = \overline {\bm x}^{\mathrm T}(n)\widetilde {\bm w}(n) + v(n)
\label{010}
\end{equation}
we have
\begin{equation}
\begin{aligned}
\mathbb{E}\left[ {\left.\frac{\overline {\bm x}(n)}{\overline {\bm x}^{\mathrm T}(n)\overline {\bm x}(n)}\left\{ {\overline e(n) - \gamma} \right\}\right|{\overline {\bm x}}(n)} \right] =&\; \mathbb{E}\left[ {\left.\frac{\overline {\bm x}(n)e(n)}{\overline {\bm x}^{\mathrm T}(n)\overline {\bm x}(n)}\right|\overline {\bm x}(n)} \right] \\ 
&- \mathbb{E}\left[ {\left.\frac{{\overline {\bm x}(n){\bm \eta^{\mathrm T}}(n)\bm{w}(n)}}{{\overline {\bm x}^{\mathrm T}(n)\overline {\bm x}(n)}}{\kern 1pt}\right|\overline {\bm x}(n)} \right] \\
&- \gamma \mathbb{E}\left[ {\left.\frac{\overline {\bm x}(n)}{\overline {\bm x}^{\mathrm T}(n)\overline {\bm x}(n)}\right|\overline {\bm x}(n)} \right] \\ 
\end{aligned}
\label{011}
\end{equation}
\begin{equation}
\begin{aligned}
\mathbb{E}\left[ {\left.\frac{\overline {\bm x}(n)e(n)}{\overline {\bm x}^{\mathrm T}(n)\overline {\bm x}(n)}\right|\overline {\bm x}(n) } \right] =&\; \mathbb{E}\left[ {\left.\frac{\overline {\bm x}(n){{\overline {\bm x}}^{\mathrm T}}(n)\widetilde {\bm w}(n)}{{{\overline {\bm x}}^{\mathrm T}}(n)\overline {\bm x}(n)}\right|\overline {\bm x}(n)} \right] \\ 
&+ \mathbb{E}\left[ {\left.\frac{\overline {\bm x}(n)v(n)}{\overline {\bm x}^{\mathrm T}(n)\overline {\bm x}(n)}\right|\overline {\bm x}(n)} \right] = 0 
\end{aligned}
\label{012}
\end{equation}
\begin{equation}
\begin{aligned}
\mathbb{E}\left[ {\left.\frac{\overline {\bm x}(n){\bm \eta^{\mathrm T}}(n)\bm {w}(n)}{\overline {\bm x}^{\mathrm T}(n)\overline {\bm x}(n)}\right|\overline {\bm x}(n)} \right] =&\; \mathbb{E}\left[ {\left.\frac{{\overline {\bm x}(n){\bm \eta^{\mathrm T}}(n)\bm {w}(n)}}{{\overline {\bm x}^{\mathrm T}}(n)\overline {\bm x}(n)}\right|\overline {\bm x}(n)} \right] \\ 
&+ \mathbb{E}\left[ {\left.\frac{\bm \eta(n){\bm \eta^{\mathrm T}}(n)\bm {w}(n)}{{\overline {\bm x}^{\mathrm T}}(n)\overline {\bm x}(n)}\right|\overline {\bm x}(n)} \right] \\ 
=&\; \sigma_\eta^2 \mathbb{E}\left[ {\left.\frac{\bm{w}(n)}{{\overline {\bm x}^{\mathrm T}}(n)\overline {\bm x}(n)}\right|{\overline {\bm x}}(n)} \right]. 
\label{013}
\end{aligned}
\end{equation}

Substituting (\ref{011})-(\ref{013}) into (\ref{008}), we have
\begin{equation}
\begin{aligned}
\mathbb{E} \left[ {\bm \xi(n) |\overline {\bm x}(n)} \right] =&\; \mu \sigma_\eta ^2\mathbb{E}\left[ {\left.\frac{\bm {w}(n)}{\overline {\bm x}^{\mathrm T}(n)\overline {\bm x}(n)}\right|\overline {\bm x}(n)} \right] \\ 
&+ \mu\gamma \mathbb{E}\left[ {\left.\frac{\overline {\bm x}(n)}{\overline {\bm x}^{\mathrm T}(n)\overline {\bm x}(n)}\right|\overline {\bm x}(n)} \right]. 
\label{014}
\end{aligned}
\end{equation}

Applying the stochastic approximation \cite{3} into (\ref{014}), the bias-compensation vector can be written as
\begin{equation}
\bm \xi(n) = \mu\sigma_\eta^2\frac{\bm{w}(n)}{\overline {\bm x}^{\mathrm T}(n)\overline {\bm x}(n)} + \mu\gamma \frac{\overline {\bm x}(n)}{\overline {\bm x}^{\mathrm T}(n)\overline {\bm x}(n)}.
\label{015}
\end{equation}

\section{Regression noise variance estimation}

In practical applications, the regression noise variance is rarely available. To overcome this shortcoming, we proposed an estimation method for regression noise variance without using the noise statistical characteristics. According to (\ref{009})-(\ref{010}), we can obtain the noise-free \emph{a priori} error
\begin{equation}
\begin{aligned}
{\overline e_f}(n) =&\; \overline e(n) - v(n) \\
=&\; {\bm x^{\mathrm T}}(n)\widetilde {\bm w}(n) - {\bm \eta^{\mathrm T}}(n)\bm{w}(n).
\label{016}
\end{aligned}
\end{equation}
Assuming that $\bm \eta(n)$, $v(n)$, $\widetilde {\bm w}(n)$ and $\bm x(n)$ are independent of each other, and using (\ref{007}), we have
\begin{equation}
\mathbb{E} \left[ {\overline e_f^2(n)} \right] = \sigma_\eta^2 \mathbb{E}\left[ {{\left\| \bm{w}(n) \right\|}^2} \right]
\label{017}
\end{equation}
\begin{equation}
\begin{aligned}
\sigma_\eta^2 = {{\mathbb{E}\left[ {\overline e_f^2(n)} \right]} \mathord{\left/
{\vphantom {{\mathbb{E}\left[ {\overline e_f^2(n)} \right] } {\mathbb{E}\left[ {\left\| \bm{w}(n) \right\|}^2 \right]}}} \right.
\kern-\nulldelimiterspace} {\mathbb{E}\left[ {{\left\| \bm{w}(n) \right\|}^2} \right]}}
\label{018}
\end{aligned}
\end{equation}
where $\left\| \cdot \right\|$ is the Euclidean norm of a vector. Note that $\mathbb{E}[\overline e_f^2(n)]$ and $\mathbb{E}[ \left\| \bm{w}(n) \right\|^2]$ are unavailable in practical. Under the ergodic assumption, the expectations of the random variables used in (\ref{017}) and (\ref{018}) can be easily estimated using moving averages as follows: 
\begin{equation}
\begin{aligned}
\mathbb{E} \left[ {\overline e_f^2(n)} \right]  = \beta \mathbb{E} \left[ {\overline e_f^2(n-1)} \right] + (1 - \beta)\overline e_f^2(n),
\label{019}
\end{aligned}
\end{equation}
\begin{equation}
\begin{aligned}
\mathbb{E} \left[ {{{\left\| \bm{w}(n) \right\|}^2}} \right] = \beta \mathbb{E}\left[ {{{\left\| \bm{w}(n-1) \right\|}^2}} \right] + (1 - \beta){\left\| \bm{w}(n) \right\|^2}
\label{020}
\end{aligned}
\end{equation}
where $0\ll\beta<1$ is a forgetting factor. Then, employing the iterative-shrinkage strategy \cite{7}, we have 
\begin{equation}
\overline e_f^2(n) = {\mathrm{sign}}(\overline e(n))\max (\left| {\overline e (n)} \right| - \tau ,0)
\label{021}
\end{equation}
where $\tau$ denotes a threshold parameter, and ${\mathrm {sign}}(\cdot)$ stands for the sign function.

\section{Simulation results}
The performance of the proposed algorithm is investigated in the context of system identification. The unknown weight vector $\bm w_o$ is randomly generated with length of $L=16$. The performance of the algorithm is estimate by the normalised misalignment (NMSD in dB)
\begin{equation}
{\mathrm {NMSD}} \triangleq 10{\mathrm {log}}_{10}\left\{\mathbb{E}\left[ {\left\| \bm{w}(n) - {\bm w_o} \right\|^2} \right]/\left\| \bm w_o \right\|^2\right\}.
\label{022}
\end{equation}

\begin{figure}[h]
	\centering
	\includegraphics[scale=0.55] {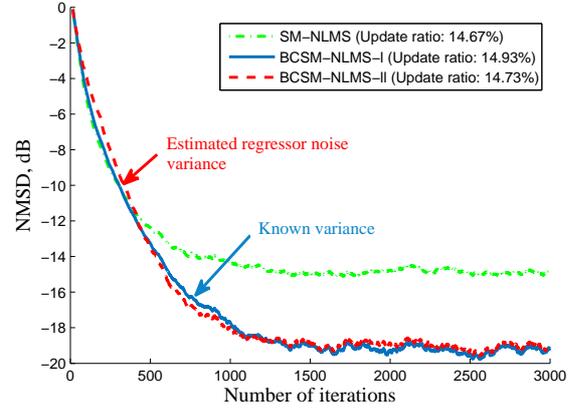}
	\caption{\emph{NMSD learning curves of the SM-NLMS, BCSM-NLMS-I and the BCSM-NLMS-II algorithms with AR(1) input (100 trials).}}
	\label{Fig02}
\end{figure}

In the first example, $\bm x(n)$ is an first-order Auto-Regressive (AR(1)) signal with a pole 0.9. A zero-mean WGN with a signal-to-noise ratio (SNR) of 10dB is added to the input, and the output is corrupted by an independent WGN with SNR=30dB. We used $\mu=0.435$ for the SM-NLMS algorithm, and $\mu=0.365$ for the BCSM-NLMS (I and II) algorithm. The parameters are set as $\beta=0.99$, $\gamma = \sqrt {5\sigma_v^2}$ and $\tau  = \sqrt {\sigma_v^2}$. As can be seen from Fig. \ref{Fig02}, the proposed algorithm outperforms the standard SM-NLMS algorithm by about 5dB. Moreover, the proposed algorithm with regression noise variance estimation method achieves similar performance to the proposed algorithm with known variance. It is interesting to note that the BCSM-NLMS II algorithm owns lower update ratio as compared with the BCSM-NLMS I algorithm.

\begin{figure}[h]
	\centering
	\includegraphics[scale=0.55] {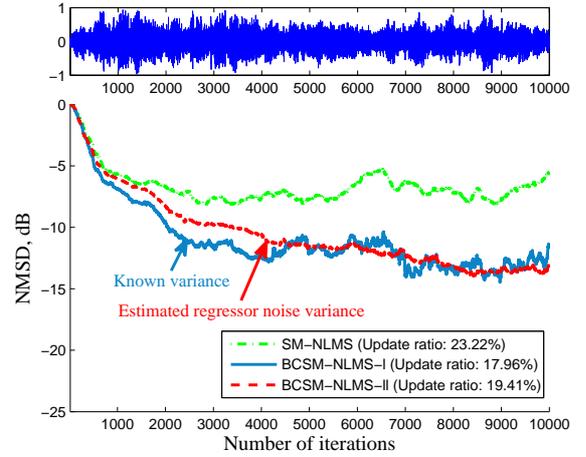}
	\caption{\emph{NMSD learning curves of the SM-NLMS, BCSM-NLMS-I and the BCSM-NLMS-II algorithms (1 trial) and speech signal.}}
	\label{Fig03}
\end{figure}

The second example is the speech input. The filter length $L$, input/output noises, $\gamma$ and $\tau$ are considered to the same as in the first example. For the SM-NLMS algorithm, $\mu=0.435$. The step sizes of the BCSM-NLMS I and II algorithms are selected to be $0.0565$ and $0.03$. Fig. \ref{Fig03} shows the NMSDs of the algorithms with speech input. We observed that the proposed algorithm yields a reduced steady-state misalignment and lower update ratio for the similar convergence rate as compared to the SM-NLMS algorithm. In particular, the performance of the proposed algorithm with regression noise variance estimation is close to that of the ideal case with known noise variance.



\section{Conclusion}
In this letter, the BCSM-NLMS algorithm for system identification with noisy inputs is proposed, which is derived from the bias-compensation approach. We further provided a novel regression noise variance estimation method, which is based the shrinkage strategy to obtain the input noise variance without using the statistics of the signals. Simulation results showed that the proposed algorithm can achieve a lower steady-state misalignment as compared with the conventional SM-NLMS algorithm.
\vskip3pt
\ack{ This work was partially supported by
	the National Science Foundation of P.R. China (Grant: 61571374, 61271340,
	61433011).}

\vskip5pt

\noindent K. Yin, H. Zhao and L. Lu (\textit{School of Electrical Engineering, School of Electrical Engineering, Southwest Jiaotong University, Chengdu, 610031, People's Republic of China.})
\vskip3pt

\noindent E-mail: hqzhao\_swjtu@126.com

\end{document}